# Mobile Data Service Adoption and Use from a Service Supply Perspective
## *An Empirical Investigation*

Krassie Petrova, Stephen G. MacDonell and Dave Parry
*School of Engineering, Computer and Mathematical Sciences, Auckland University of Technology,
55 Wellesley St. E., Auckland 1010, New Zealand*

**Abstract**
*The paper presents the findings of an empirical study of the views of a selection of mobile data service (MDS) supply chain participants about anticipated MDS customer requirements and expectations, and about the MDS environment. Applying an inductive thematic analysis approach, the study data are first represented as a thematic map; the thematic map is then used to formulate propositions that contribute an MDS supplier perspective to models investigating MDS customer adoption and use.*

## 1. INTRODUCTION AND BACKGROUND

Mobile data services (MDS) are designed, developed, provided and consumed within a mobile service ecosystem in which MDS suppliers and MDS customers interact and create service value Becker et al., (2012), (Basole and Karla, 2012; Dennehy and Sammon, 2015). The MDS ecosystem has a complex structure. On the one side, it includes mobile technology providers (e.g., mobile network operators – MNOs and mobile device vendors), and MDS developers and providers such as banks offering mobile banking services, and mobile application ("app") developers distributing their apps through mobile app marketplaces (Petrova and MacDonell, 2010; Ryu et al., 2014). On the other side, the MDS ecosystem comprises a highly heterogeneous customer space, with customer segments determined by both demographic and attitudinal factors (Floh et al., 2014).

**1.1 MDS Customers**

The extant literature is rich in empirical studies of MDS adoption and use from a customer perspective. MDS customer decisions to adopt, use , and continue to use an MDS have been studied extensively applying existing and well-validated technology adoption models (Sanakulov and Karjaluoto, 2015; Ovčjak et al., 2015). However, a number of more recent studies consider MDS adoption from a service, rather than from technology perspective (Thong et al., 2011). The approach is based on the premise that mobile technology use is subsumed by MDS use (Becker et al., 2012) as the technology users conceptualized in traditional information systems research have become instead "service consumers" (Tuunanen et al., 2010). Consequently, customer perceptions about service value and service quality have been studied as important factors influencing customer intention to adopt, use, and continue to use MDS (Kuo et al., 2009; Tojib and Tsarenko, 2012; Kim et al., 2013).

Perceived service value reflects customer perceptions regarding the overall utility of a service based on the customer's assessment of the perceived service benefits and disadvantages, and perceived service-associated "sacrifices" (cost of acquisition) (Schilke and Wirtz, 2012). Therefore, in order to ascertain a particular service's potential to generate customer demand (leading to actual service use), it is important to understand what specific value the different customers of a particular MDS attach to it (Bouwman et al., 2009).

According to Bina et al. (2007), customers value mobile services that enhance both the utilitarian and the hedonistic aspects of the their everyday life style. However, it is also observed that demographically different customer groups may have different preferences with respect to MDS type, content and interface (Lee et al., 2009; Constantiou et al., 2007), and that the demographic segments are relatively narrow (Oh et al., 2008).

Service use, in particular, may be also significantly influenced by perceived service quality. Akter et al., (2013) define perceived mobile service quality as an overall judgment about a service's "excellence". Customer perceptions about service quality (i.e., perceptions about how the service performs with respect to content, interface, navigation, and visual appeal) have been found to influence both perceived service value, and perceived post-use satisfaction (Kuo et al., 2009).

Overall, throughout the process of MDS adoption and continued use, customers make decisions significantly influenced by their perceptions about service value, and by their experiences with service performance and quality. Through actions based on these decisions (i.e., considering a service, using it occasionally or regularly, or discontinuing use) MDS customers interact with MDS suppliers; MDS customers may even have an impact on the regulatory environment through their participation in consumer groups (Camponovo and Pigneur, 2003).

### 1.2 MDS Suppliers

There has been limited work to date that considers MDS adoption and use from an MDS supplier perspective. The opinions of MDS providers with respect to MDS success/failure are investigated in (Carlsson and Walden, 2002; Scornavacca and McKenzie, 2007). Further examples include service- specific explorations such as Okazaki's (2005)'s study of the perceptions of senior executives about using mobile technology as an advertising channel, analyses of merchants' attitudes towards mPayment (Mallat and Tuunainen, 2008; Hayashi and Bradford, 2014), and a study of managers' perceptions about SMS-based marketing (Li and McQueen, 2008). A comparative analysis of the views of MDS suppliers and customers is provided in Xinyan et al., (2009) and Akesson's (2007) investigations of mPayment and mobile media content provision, respectively.

The reviewed studies indicate that perceptions about customer needs and behaviour, and perceptions about the context in which the service is offered influence MDS supplier decisions about investment in service development. In addition, the findings of the last two studies indicate that customer and service provider opinions and views with respect to the importance and role of MDS adoption factors may differ; corroborating results (from a study of mobile communication services) are reported in (Abu-El Samen et al., 2013). The findings form the literature provide support for the assumption that underpins the research presented here, namely, that MDS supplier perceptions about customer demand for MDS influence the MDS development and provision.

More specifically, it is contended that MDS supplier perceptions about the targeted customer group requirements and expectations affect the MDS value proposition with respect to service functionality, design, and pricing model, and that perceived customer demand represents MDS supplier knowledge and understanding of both existing and potential customers' needs for MDS, their quality of service expectations and daily lifestyle related requirements, and the relevant service and regulatory environment.

### 1.3 Research Aim and Questions

Based on the analyses above it is suggested that a better understanding of MDS supplier perceptions about customers may contribute to a better understanding of supplier–customer interactions in the MDS ecosystem, including the development of the MDS value proposition and its acceptance by customers. The research presented here aims to propose an MDS supplier perspective on MDS customer adoption, as supported by the outcomes of an empirical investigation of the views of MDS suppliers about customer demand for MDS. Three specific research questions guide the empirical investigation:

1. What are MDS supplier views about customer expectations, requirements, and attitude drivers?;
2. What are MDS supplier views about the value of customer mobility support features of MDS?
3. What are MDS supplier views about the mobile service supply and regulatory environments?

As asserted earlier in (Petrova and MacDonell, 2010), an MDS supplier perspective on what customers need, want, expect, and get from MDS may complement prior studies that mainly looked at MDS adoption and use from a customer perspective. The current study that contributes a set of propositions that can be used to extend MDS adoption and use models.

The rest of the paper is organised as follows: the next section outlines and justifies the research approach, and describes the data gathering, coding and analysis processes. Sections 3 and 4 present and discuss the findings. Section 5 highlights the study contributions and limitations, and suggests directions for further research.

## 2. RESEARCH METHOD

The study aims to achieve an understanding of a phenomenon (i.e., how MDS supplier perceptions about customers may influence MDS adoption) from the view point of the research participants; therefore, it was considered appropriate to apply a qualitative investigating approach that followed a "from the ground up" (Creswell, 2007) logic. As the MDS supplier space includes different types of organisations, the adopted research method was collective case study (Onwuegbuzie and Leech, 2007); because of the exploratory nature of the investigation the data were analysed inductively (Patton, 2002). An inductive thematic analysis process (Braun and Clarke, 2006) was developed; applied systematically and interactively, it allowed to identify the patterns and the themes emerging across the data set. Thematic networks (Attride- Stirling, 2001) were constructed and used to organise the emerging themes in a thematic map, and interpret the findings further.

### 2.1 Study Setting

The empirical investigation took place during the period 2010-11 in Bulgaria. At the time of the study the mobile telecommunications sector included three MNOs (subscriber penetration rate of 141% - http://en.wikipedia.org/wiki/List_of_mobile_network_ope rators_of_Europe).

The MNOs had already started offering MDS such as mobile payment (mPayment) (http://paper.standartnews.com/en/article.php?d=200 7-08-02&article=5980) while a number of software houses had engaged in developing mobile games and other mobile entertainment applications (http://gdsbulgaria.com/en/Studios). The regulatory and legislative infrastructure included the provisions of the Communications Regulation Commission (Pook, 2008) and specific pieces of legislation such as the Law on electronic commerce and the Law of fund transfers, and electronic payment instruments and payment systems.

## 2.2 Research Instrument and Sample

Semi-structured interviews (Myers, 2009) were adopted as the primary method for data gathering, chosen because of the need for flexibility while talking to a diverse range of participants. The questionnaire was tested in a pilot study; the work of Tilson et al., (2008) provided a useful reference. The final version contained five background questions and 12 knowledge/opinion/value judgment questions about MDS value proposition, anticipated customer attitude and MDS acceptance behaviour, and the service and regulatory environment.

The study participants were recruited from amongst the employees of companies and organizations involved in MDS design, development and provision. To ensure that participants were able to offer well-informed opinions, the study sought to recruit individuals with significant expertise who were both knowledgeable about the area of the investigation, and were involved in decision making at their respective place of employment.

A total of 52 individuals matching the profile above, from 13 organizations, were issued invitations; eventually, 12 individuals from eight organizations accepted to be interviewed. Nine participants were employed by "large" companies (staff above 250) while three worked for "small" ones (staff between 50 and 250, applying the European Union definitions for small and medium enterprises). It occurred that five participants (MD1- MD5) were involved predominately in roles related to software development and MDS design, while the remaining seven participants (MP1-MP7) were more closely associated with MDS content development, provision, and aggregation. The sample size was deemed adequate for a single case study; it exceeded significantly the four to five participants recommended by Creswell (2007) but allowed to represent the different MDS supplier types.

## 2.3 Data Coding and Analysis

The interviews were transcribed (six transcripts were translated from Bulgarian into English) and stored electronically. The coding started with developing a set of deductive codes based on the research topic and applying them in order to gain an initial understanding of the data. Next the data were interpreted and coded inductively employing in vivo coding (Saldaña, 2012). A rigorous iterative coding protocol was developed and followed in order to preserve traceable links between data, interpreted data meanings, and data codes. The coded data were organised into a multilevel data dictionary.

Ultimately, 413 data meanings were extracted, interpreted and assigned a code; the code labels and definitions were kept close to the data meanings.

Similarly coded data were grouped together under a "super code". The final version of the data dictionary contained data supporting 99 super codes grouped in five mutually exclusive groups of related coded meanings (categories): (i) CUSTOMERS (perceptions about customer space characteristics including attitudes, behaviours, requirements, and expectations), (ii) SERVICE SUPPLY & DEMAND (perceptions about MDS supply space characteristics including service value and viability), (iii) TECHNOLOGY (perceptions about technology opportunities and limitations), (iv) REGULATORY ENVIRONMENT (perceptions about the regulatory environment), and (v) UNCERTAINTY (what participants felt uncertain about).

Table 1: Emerging and organising themes.

| Emerging themes | Organizing Themes |
|---|---|
| Difficult customers; Customer segmentation | Customers differ |
| Attractive services; User friendly services | Customers require |
| Need for service; Service value | Customers expect |
| Personal goals; Free services | Customers prefer |
| Optimistic providers; Service innovation; Reg. environment opportunistic | Opportunities and challenges |
| Operators as a barrier; Operators threatened | Barriers |

## 3. FINDINGS

The coded data were iteratively and systematically re-examined in order to identify and define an initial set of emerging themes. The emerging themes were organised into a thematic map, and the key points made by the participants were extracted.

### 3.1 Emerging Themes

The emerging themes were identified through pattern coding, searching for semantically related super codes; the super codes were methodically considered with respect to forming discernible patterns that may be interpreted as a coherent theme. The set of semantic relationships (adapted from Gibson and Brown, 2009) included "Associated with", "Aspect of", "Cause of/Result of", "Contrast with", and "Attribute of".

As shown in Table 1 (first column), a total of 13 themes emerged; the themes did not overlap across the data, as each super code could be associated with one theme only. Brief theme descriptions are provided in the Appendix; the labels of the super code associated with each theme highlight the theme's key characteristics.

At the next step of the analysis the data supporting the emerging themes were searched for similarities or shared ideas that could be used to interconnect the themes. Six similarity clusters were identified and used to define six higher level "organizing" themes (Table 1, second column).

Finally, the supporting data were examined again in order to determine the top-level theme grouping - the overarching "global" themes epitomizing the key points, or main meanings of the data. Figure 1 shows the overall outcome of the analysis: a thematic map representing the original data set as two related thematic networks, each centred on one of the two global themes described below; the illustrating data codes are referenced to the cited participants.

### 3.2 Global Theme "Customers Demand"

Global theme "Customers demand" pertains to the characteristics and behaviour of potential and actual MDS customers as perceived by the study participants, and the effect of these perceptions on MDS supply. Organising theme "Customers differ" (for data quotes, see Table 2) reveals that MDS suppliers perceive customers as distrustful and conservative in their attitude, while having a wide range of different requirements and expectations. Therefore, the customer market is perceived as very segmented and hard to satisfy; the situation is exacerbated by a certain lack of sufficient understanding of what customers really want. From an MDS supplier perspective, the mobile service industry may not be offering the services customers need, or expect, i.e., there may exist a misalignment between MDS supply and MDS demand.

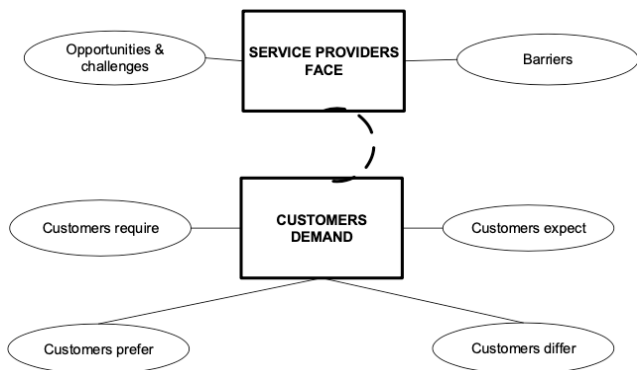

Figure 1: Thematic map.

Further insights into MDS supplier perceptions about MDS customer requirements and expectations are offered by the organising themes "Customers require", "Customers expect" and "Customers prefer" (data quotes related to each theme are provided in Tables 3, 4 and 5). First, the data indicate that according to MDS suppliers, customers would be interested in new services that were not just innovative but also engaging, easy to access and use, and well supported. From an MDS supplier perspective, significant effort is required in order to overcome the inherent technology limitations and develop MDS attractive enough to compete with similar services developed for personal computers.

Table 2: "Customers differ" - data quotes.

| | |
|---|---|
| MD3 | "security first…the majority of users don't easily rely on innovations"; |
| MP1 | "Inertia of older consumers, expressed in fear and resistance against innovations and developments"; |
| MD5 | "it is difficult to persuade customers to break with the old routines and influence them towards adopting new innovative products if the need to do so is not urgent"; |
| MD1 | "Yes, there are [segments] and their expectations are different"; |
| MD3 | "Consumers can be divided into groups of expectations …some seek security and usability, other entertainment, facility, etc."; |
| | "to offer something new and better… not an easy task…there are no clear criteria exactly what the market wants… mobile applications are relatively new, but despite this… the 'saturation' effect is visible"; |
| MD4 | "Development of new services is going ahead of demand". |

Second, MDS supplier perceive customers as expecting services of high value and with a clear value proposition that balances quality and cost. Perceived customer expectations include well designed services that meet specific customer needs, make innovative use of connectivity, offer compatibility across devices and platforms, provide privacy protection, and are reasonably priced. Furthermore, customer judgement is perceived as influenced by external factors such as peer opinion, and the [lack] of sufficient knowledge about MDS. From an MDS supplier perspective, the challenge is to make sure that customers understand the MDS value proposition and associate it with a perceived service need.

Table 3: "Customers require" - data quotes.

| | |
|---|---|
| MD2 | "Interesting ideas that would motivate people to use new development"; |
| MP5 MD3 MP7 | "The service has to be 'modern': "contribute to a richer user experience"; "user-friendly…easy and fast accessibility and support 24 hours a day"; |
| MD4 | "Developers… are restricted by the limited resources of the mobile device… so, with much less options an application has to be developed that does not defer drastically to those, made for PCs"; |
| MP3 | "Innovations are needed so that the application is [made] attractive". |

Table 4: "Customers expect" - data quotes.

| | |
|---|---|
| MP2 | "added value… has direct impact on the customer"; |
| MP7 | "Customer attitude is affected [by] does it add value to the service"; "Internet banking now works well and customers want it on their mobile phones"; |
| MD5 | "for health apps, customers are …willing to pay more and price is not such a big issue"; |
| MP3 | "…offer connection to any kind of other devices – TVs, cars to be operated via mobile phone"; |
| | "good mobile software performance"; "information about the availability of such a service and how to use it"; |
| MP2 | "Compatibility with various OS as Android, Windows Mobile"; |
| MP4 | "safety of the personal information and the user's data"; |
| MD1 MP5 | "price and quality"; "Can only be convinced by opinions …friends who have good impressions"; Becoming aware of the need for a certain service…implied need through advertisement". |

Specific MDS service characteristics that (according to MDS suppliers) customers perceive as adding value to MDS include meeting personal requirements for convenience and availability (e.g., saving time), being able to choose from a range of services, and opportunities for customer involvement. Furthermore, customers are perceived as preferring functionality over feature overload, with service availability and high network performance quality the other top preferences. From an MDS supplier perspective, the viability of developing and providing value-adding MDS is affected by customer attitude towards paying for MDS. Here, participants differed in their views: according to some, customers would be willing to pay for a service they felt was valuable while others saw customers as not prepared to pay "too much" for a service anyway.

The uncertainty affects MDS supplier decisions about how to approach service design and implementation – to invest significantly in MDS that satisfy customer requirements and expectations about service value, or to offer less valuable, low cost service versions that are more likely to be accepted and may help create a critical customer mass. However, participants expressed reservations about the acceptance of free MDS: they perceived potential MDS customers as discerning and likely to consider a free MDS as one of a lesser value.

Table 5: "Customers prefer" – data quotes.

| | |
|---|---|
| MD3 | "the convenience to be able to do whatever you want, whenever and wherever you want - something very important because it saves time"; |
| MP6 | "Innovation is very important in this sector; the customer has the choice how to get something done"; |
| MP2 | "customer to be able to control, monitor and act pro-actively…the product features and the product flexibility always prevail vs. the 'fashion design'…24/7 service availability and support"; |
| MP1 | "Accessibility at any time and from anywhere to information resources as well as speed in obtaining information". |
| MD5 | "What really matters is the value that the mobile product brings and how desired the solution is"; |
| MP7 | "Private users usually are not ready to pay a considerable price and the cost-value relation is an especially important part of their motivation to purchase the product"; |
| MD4 | "I firmly believe that a given free product can give much more profit with its popularity, than a product that is paid and because of this – less used/less known"; |
| MD2 | "free applications attract the interest of people, but if they are not well made and sufficiently functional, as is usually the case with free stuff, the user would rather not use that application or would consider buying the paid version, which will have a much better good maintenance". |

**3.3 Global Theme "Service Providers Face"**

In global theme "Service providers face" participants talk about the dynamics of the MDS supply sector and the MDS supply environment. The focal points are the opportunities and challenges in developing and offering new MDS, the barriers faced by MDS providers, and the role of MNOs.

According to organizing theme "Opportunities and challenges" (for data quotes, see Table 6), the increased affordability of smart phones and the relatively supportive regulatory environment create opportunities for MDS development (e.g., services targeting niche areas of customer needs, services that complement existing ones). The MDS supply environment is dynamic and competitive, therefore, profitability should be considered from a strategic perspective. Both being innovative and staying ahead of competitors, and following successful innovators are viable options, provided that the service offers a satisfactory customer experience.

However, MDS development faces two major challenges. The first challenge relates to the still considerable inherent limitations of the mobile technology. While these limitations can be overcome by innovative service design, MDS developers need a better knowledge of the targeted customer segment (as discussed in the first global theme) in order to align the level of service interface sophistication with the level of customer comfortableness with the technology. The second challenge is specific to the case context. The relatively small local market may not be conducive to developing innovative services as there exists an uncertainty about service viability exacerbated by the segmented customer market and customer attitude towards paid MDS (also discussed in the first global theme).

Table 6: "Opportunities and challenges" – data quotes.

| | |
|---|---|
| MD4 | "with…smartphones…becoming cheaper, mobile technologies will get more attractive"; |
| MD5 | "the regulatory environment is relatively supportive, except for … private data abuse in terms of location based services and private person location information"; |
| MP3 | "a forthcoming boom …is to be expected"; "it is often easy for the developers of a mobile application to fill a 'niche' in the market"; |
| MP7 | "[good] user experience always brings more benefit and affects the customer's satisfaction"; |
| MP2 | "where a product is similar to other offered by other providers, its competitiveness requires more added value and more specific and eloquent advantages"; |
| MP7 | |
| MP7 | "In Bulgaria our mobile service is among the first ones which makes it especially valuable for customers"; |
| MP6 | "as companies are striving to be the best, they develop services not orientated mainly towards …profit, but … important for the image"; |
| MD4 | "Having in mind the limitations of mobile devices [that…still do exist], it is important to know who exactly the users… will be…the …service has to be in precise conformity with their technical knowledge and potentialities"; |
| MP7 | "The Bulgarian mobile market is small in comparison to bigger countries with a larger number of users… new solutions are offered relatively late here". |

According to organizing theme "Barriers" (for data quotes, see Table 7), mobile operators are not too interested in providing and/or supporting MDS and prefer to focus instead on their traditional services. The still relatively high mobile data cost represents a significant barrier to MDS development and provision as it affects negatively MDS customer adoption and MDS viability. However, the environment is becoming more competitive; fearing revenue loss, MNOs may become more interested in MDS development and provision.

From a MNO's point of view mobile operators should sustain their leadership role in the MDS supply sector; an inherent impediment to the implementation of innovative ideas is the lack the flexibility caused by organizational complexity and slow internal processes This increases the risk associated with MNOs involvement in MDS development and provision.

Table 7: "Barriers" – data quotes.

| | |
|---|---|
| MD4 | "Operators are rather in the way of mobile applications distribution... internet traffic has the lowest priority in mobile devices in comparison with telephony... Who would want an application that would work only if it had free resources not used for telephony...Mobile internet prices...are making the use of applications expensive and thus unattractive...the investment in mobile applications is still not very profitable"; |
| MD5 | "The highest profit is made by standard services...because of competition prices are falling down and the operator has to be innovative and constantly work on its services"; |
| MP2 | "facing the treat to lose the customers loyalty and become only the transport link to the end-user services...the telecom operators should have the leading role having in mind that the connectivity is important"; |
| MP4 | "new services may be useful as well as a threat and that is why they need to have a role in these services"; |
| MP7 | "The time needed for planning and designing a mobile application is too long...the service is no longer attractive or needed in the time of its launching ... there are other similar services". |

## 4. DISCUSSION

The findings of the data analysis highlight the key points made by participants. We explore them further by addressing the research questions and developing propositions for a follow up empirical investigation of MDS adoption and use. With respect to positioning the propositions within the extant literature on MDS customer adoption and use we refer both to empirical work, and to work that critically reviews prior results. As already mentioned the literature on MDS supplier perceptions with respect to MDS adoption is scarce; therefore, we compare our findings to the outcomes of Shieh et al.'s (2014) investigation. Shieh et al. examined the relevant literature, extracted a range of factors found to be affecting MDS customer adoption, and asked a group of mobile telecommunication experts, knowledgeable about MDS, to rank the factors according to their perceived importance; the authors report on factors ranked from one (most important) to ten only.

The key points made in first global theme allow to address explicitly the first two research questions (i.e., MDS supplier views about customer expectations, requirements, and attitude drivers, and about the value of customer mobility support features of MDS). Overall the data suggest that according to MDS suppliers, customer attitude towards MDS adoption and use is driven by perceptions of service value, based on customers' considerations about how much they need a particular service, and how well the service is expected to perform (or has performed).

In prior empirical research about customer adoption of MDS perceived value is considered as directly influencing customer intention to use and actual use of MDS. Two dimensions - mobility support and mobile technology performance related features are identified in (Johansson and Andersson, 2015) and in (Al-Debei and Al-Lozi, 2014; Ervasti, 2013), respectively.

Our data indicate that customers are perceived as not differentiating between the technologies used rather considering mobile services as a specific new type of online (Internet) services. Furthermore, customers are perceived as evaluating the service proposition according to their needs; while they would expect services to take advantage of innovative features such as anywhere/any time access customers assess services primarily with respect to meeting their specific requirements and personal goals. These findings are similar to Shieh et al., (2014) – in their study the factors related to mobility support ("service accessibility" and "real- timeliness") are at the bottom of the ranking table (in 8th and 9th position, respectively).

While perceived service need varies with individuals' lifestyles, and also according to the characteristics of specific customer market segments, the dimensions of perceived service quality are more uniform and include service delivery quality (i.e., service availability and support), and service performance quality (i. e., service functionality and service design, including the user-friendliness of the interface). Partial corroboration of these findings can be found in (Shieh et al., 2014) where "network coverage", and "comprehensive customer service" are ranked 6th, and 4th, respectively. However the constructs service design and service functionality are not explicitly included in the models reviewed.

We formulate the following propositions that reflect MDS supplier views on customer attitudes and expectations with respect to service value:

**P1:** Perceived service value influences positively attitude towards MDS adoption.

**P1.1:** Perceived service need influences positively perceived service value.

**P1.2:** Perceived mobility support influences positively perceived service need.

Empirical research about customer adoption and use of MDS also includes system quality (related to performance quality) as an indirect factor influencing (Ovčjak et al., 2015). In (Shieh et al., 2014) two related factors – signal quality and transmission speed, are ranked 2nd and 7th, respectively. We formulate the following propositions that reflect MDS supplier views with respect to service quality:

**P2.1:** Perceived service delivery quality influences positively perceived service value.

**P2.2:** Perceived service performance quality influences positively service value.

It was shown that two opposing views emerge with respect to customer acceptance of service cost: according to some, customers would be prepared to accept the service cost if the service meets their expectations and requirements while according to others, customers will always prefer a low cost/free service to a paid one regardless of the perceived value. Furthermore, in the case of MDS that compete with similar services offered through other channels, customers have a choice, and may prefer a non-mobile version of the service based on cost comparison.

Extant research has been inconclusive about the role of perceived cost in MDS adoption and use. For example,

perceived cost is not explicitly included in the model proposed by Troshani and Hill (2008) which is based on a synthesis of prior research; in the review by Sanakulov and Karjaluoto (2015), while perceived cost is not found to influence MDS adoption and use in mobile banking and mobile learning studies, it is found to affect MDS adoption and use in studies that do not focus on particular services. Furthermore, in (Ovčjak et al., 2015) perceived cost is part of the proposed conceptual model for the adoption of mobile information services but is not a factor in the mobile entertainment, and mobile transaction conceptual adoption models. However, the participants in (Shieh et al., 2014) consider cost important ("handset prices and transmission fees" are ranked 4th). We formulate the following propositions that reflect MDS supplier views on customer attitude towards MDS cost:

**P3.1:** Perceived service cost influences perceived service value.

**P3.2:** Perceived service cost influences attitude towards MDS adoption.

Global theme "Service providers face" provides insights into the third research question (i.e., MDS supplier views about the mobile service supply and regulatory environment). We look at how MDS supplier views about the service and regulatory environment relate to the propositions developed earlier, and allow to formulate new propositions.

First, participants are well aware of the complexity of the customer market and the need to deal with the high customer expectations but are somewhat uncertain about what customers really want. However participants are relatively optimistic about the future of MDS seeing it as driven by rapid technological progress that generates need for new services. These views provide support for proposition **P1.1** developed earlier.

Second, an MDS supply environment related factor affecting MDS adoption is the high network access cost (supporting propositions **P3.1** and **P3.2**). Mobile technology limitations also play a role as they make it difficult to achieve service design comparable (in terms of design quality) with other online services (supporting proposition **P2.2**).

Third, participants consider MDS viable in the long term and see market opportunities such as developing very specialized services (supporting proposition **P1**) and taking advantage of innovative technology opportunities to distribute services and reach the target customer market (supporting proposition **P2.1**).

Furthermore, participants identify offering an innovative service ahead of other competitors as an opportunity to develop a financially viable service. While customers' personal attitude towards innovation is considered in empirical work, e.g., in mobile banking adoption studies (Shaikh and Karjaluoto, 2015); similarly personal innovativeness is included in the conceptual model for mobile transaction services adoption in (Ovčjak et al., 2015), but not in their mobile information and mobile entertainment conceptual adoption models. We formulate the following proposition reflecting MDS supplier views about the role of innovation as a service design characteristic:

**P2.3:** Innovative service design and functionality influences perceived service need.

Finally, mobile network operators are seen as reluctant to support MDS; the implications related to perceived quality of service delivery are captured by proposition **P2.2**. The regulatory environment is seen as mature (with some applicable legislation already in place), and relatively supportive, or at least not presenting any significant obstacles to MDS development and deployment. It may be inferred that according to participants there is no need for any specific further development in the regulatory environment space.

The propositions developed above contribute towards the development of an MDS customer adoption model that takes into account the views of MDS suppliers. The introduction of a new variable (service need). in proposition **P1**, and service specific variables (service delivery quality as determined by perceived service availability and support, and service performance quality as determined by perceived service design and functionality in propositions **P2.1** and **P2.2**, respectively) is in line with suggestions that there is a need for new constructs and relationships when studying the adoption and use of advanced technologies (Sanakulov and Karjaluoto, 2015).

Propositions **P2.1** and **P2.2** can be considered as a specific instance of the relationship between quality and value as synthesised in (Cronin et al., 2000); proposition **P2.3** refers to service innovativeness as a catalyst to service adoption through its influence on perceived service need and is, therefore, a facilitating condition (Rao and Troshani, 2007). Last, the dual approach to considering perceived service cost as both a value forming factor (proposition **P3.1**), and an attitude driver (proposition **P3.2**), may lead to a better explanation of its role specifically with respect to MDS adoption.

## 5. CONCLUSIONS

In this paper we present the findings of an empirical study that looks at MDS provision and adoption from the view point MDS suppliers, based on a single case study. The study contributes to the understanding of how MDS supply stakeholders develop the related service value proposition – a direction for further research suggested by Shaikh and Karjaluoto (2015). More specifically, this study develops propositions that may add an MDS supply perspective to existing MDS customer adoption and use models by considering customer perceived service need, and delivery and performance quality as major drivers of customer perceived service value.

With respect to practical implications, the findings of the study imply that MDS need to be developed with one or more specific customer segments' existing or potential needs in mind, and with focus on overall service performance, customer experience, and customer engagement. Second, MDS suppliers should seek partnerships and collaborations with infrastructure providers (e.g., MNOs) in order to develop more affordable services, and increase customer service awareness.

The study has two major limitations. First, empirical data were collected in 2010; however there is evidence to suggest that the context of the study has not changed significantly compared to 2010 (e.g., Kraleva et al., 2016, Otuzbirov and Aleksiev, 2015); therefore, the findings may still be relevant. Second, the study is set in a specific country context, and draws inferences from the interpretation and inductive analysis of qualitative data; therefore, extending the findings to other contexts and further theory building may need conducting similar investigations in different settings, based on the same theoretical assumptions and applying the methodology developed and tested, including the data coding scheme.

Other directions for further research include empirically validating the propositions, studying factors that influence MDS supplier perceptions, and investigating how MDS suppliers may see customers as participants in the process of innovative service value co-creation, for example, adapting the consumer value co-creation framework proposed in (Tuunanen et al., 2010). Finally, it would be of interest to investigate how the prevailing MDS business models facilitate the development of an acceptable MDS value proposition, by integrating the findings of this study with existing frameworks e.g., Sharma and Gutiérrez's (2010) mobile commerce business model evaluation framework.

# APPENDIX

# Emerging Themes

| **"Difficult customers"**  *Customers difficult to satisfy, conservative.* |
|---|
| Customer market difficult; Customers conservative/inertia; Customers distrustful of innovation; Customers distrustful of phones; Customers prefer well known services; Segmentation by age – young customers; Segmentation by attitude to innovation; Expectations about quality high; Expectations difficult to meet; Expectations for choice of services; Service to surpass existing ones; Services not different from existing ones; Lack of knowledge about customers. |
| **Customer segmentation**  *Customer market very segmented* |
| Segmentation by specificity of requirements; Segmentation by age; Segmentation by self-efficacy; Segmentation by socio-economic status; Segmentation is multidimensional; Customers do not mix entertainment and serious business; Decision influenced by cost – not; Decision influenced by cost; Narrow customer base. |
| **Attractive services**  *Appealing design and innovative features attract customers* |
| Expectations for appealing service design; Expectations for rich experience; Services that are attractive to customers; Free trial increases popularity; Free services attractive if modelled on successful paid ones; Paid services less attractive; Customer motivation needed to stimulate development. |
| **Free services**  *Free services draw customer attention* |
| Decision influenced by cost ongoing; Decision influenced by service affordability; Expectations for low service cost; Free services valued; Low cost service valued; Free services not reliable; High service cost due to high data cost; Free services profitable if very popular; Service with some free functions may be successful; Cheap applications already available. |
| **Need for service**  *Services are viable if customers see them as meeting their specific needs* |
| Decision influenced by how much a service is needed; Decision influenced by marketing; Service needs to be meeting a need; Service not useful; Service not meeting a need not valued; Attractive use scenarios exist. |
| **User friendly services**  *Customers require services to be 'friendly* |
| Customers require services to be friendly; Decision influenced by ease of use; Service needs to be easy to use; Usability valued. |
| **Personal goals**  *Customers choose services based on personal goals* |
| Service needs to focus on personal mobility; Service needs to meet personal goals; Service saturation; Anytime/anywhere services valued; Customer empowerment; Services matching personal lifestyle valued. |
| **Service value**  *Customers look for service value* |
| Decision influenced by added value; Decision influenced by comparison; Decision influenced by compatibility; Decision influenced by cost-effectiveness; Decision influenced by service quality; Decision influenced by social norm. Expectations for high service performance; Expectations for service value; Expectations for support; Service needs to be convenient; Connection with other devices valued; Paid services with support valued; Security fears. |
| **Optimistic providers**  *Service providers believe in the future of mobiles services* |
| Current use; Need for entertainment services; Changing market; Competition; Environment; First on the market; User experience; Viability potential; Successful models exist. |
| **Service innovation**  *Mobile technology offers potential that can be captured through innovative approaches* |
| Limitations due to device design; Technology not available yet; Technology limits architecture; Service needs to be technologically implementable; Potential opportunities; Opportunities offered by device design; Opportunities to distribute services; Opportunities to support customers; Uncertainty about technology |
| **Regulatory environment opportunistic**  *Regulatory environment not restrictive, offers opportunities* |
| Regulations exist that are also applicable; No regulations; Regulation needed – some; Regulatory environment - lack of awareness; Regulatory environment supportive; Regulatory environment moderately supportive; Regulatory environment changing. |

| **Operators as a barrier** |
| :---: |
| *Mobile network operators act as a barrier to mobile service development* |
| Operators as a barrier to service; High investment cost; Lack of operator support for development; Low quality of service due to lack of operator support. |
| **Operators threatened** |
| *Mobile network operators are facing a threat* |
| Loosing competitive advantage; [Other] Players; Uncertainty about MNOs. |